# PLANCK / LFI: AN ADVANCED MULTI-BEAM HIGH PERFORMANCE MM-WAVE OPTICS FOR SPACE APPLICATIONS

F. Villa[1], M. Sandri[1], M. Bersanelli[2], R.C. Butler[1], N. Mandolesi[1], A. Mennella[3], J. Marti-Canales[4], J. Tauber[4]

on behalf of LFI consortium

[1]*IASF/CNR – Sezione di Bologna*
*Via P.Gobetti, 101*
*I-40129 – Bologna, Italy*
*Email: villa@bo.iasf.cnr.it*

[2]*Università degli Studi di Milano*
*Via Celoria, 16*
*I-20133 – Milano, Italy*
*Email: marco@mi.iasf.cnr.it*

[3]*IASF/CNR – Sezione di Milano*
*Via Bassini, 15*
*I-20133 Milano, Italy*
*Email: daniele@mi.iasf.cnr.it*

[4]*ESA – ESTEC*
*PO Box 299*
*2200 AG Noordwijk, The Netherlands*
*Email: Javier.Marti.Canales@esa.int*

## INTRODUCTION

The Low Frequency Instrument (LFI) is one of the two instruments onboard the ESA PLANCK satellite foreseen to be launched in 2007. The LFI will image the Cosmic Microwave Background (CMB) anisotropies and the polarization in four different bands, with an unprecedented combination of sky coverage, calibration accuracy, control of systematic errors, and sensitivity. LFI is coupled to the PLANCK Telescope by an array of 23 high performance dual profiled corrugated feed horns. The dual reflector off axis-design of the 1.5 meter projected aperture telescope is the most advanced optical design ever conceived for accommodating large multi-beam and multi-frequency focal plane units.
The location and the design of the feed horns, for both instruments, have been optimised in order to maintain good beam symmetry, even far from the optical axis, and an excellent straylight rejection. In this paper we describe the LFI instrument with emphasis on its optical performance and coupling with the PLANCK telescope.

## THE PLANCK MISSION

PLANCK is the third European Space Agency medium-sized (M3) mission of the `Horizons 2000' science programme, and is one of the three missions (together with HERSCHEL and EDDINGTON) of the Astrophysics / Group 2 of the new ESA science programme "Cosmic Vision 2020". With its 1.5 meter passively cooled telescope and the two multi–beam instruments (the Low Frequency Instrument, LFI, covering the 30–100 GHz range; and the High Frequency Instrument, HFI, in the 100-850 GHz range) both located at its focus, PLANCK will simultaneously observe the sky in nine frequency bands across the maximum of the CMB planckian spectrum [1] [2]. The mission will produce full-sky maps of the anisotropies of the CMB with unprecedented sensitivity (few μK per pixel) and an angular resolution between 5' and 33' depending on frequency (typically 10 arcmin for the 100 GHz channel). Moreover, it is expected that PLANCK will measure the statistics of the predicted polarized E– component of the CMB at sub–degree angular scales with excellent sensitivity, especially in the regions near the ecliptic poles. The multi-frequency maps will represent a major source of information relevant to several cosmological and astrophysical issues, such as testing theories of the early universe and the origin of cosmic structure, the knowledge of the cosmological parameters, the measurement of the Sunyaev–Zel'dovich effect, as well as several astrophysical aspects such as studies of discrete sources and galaxy diffuse emissions.

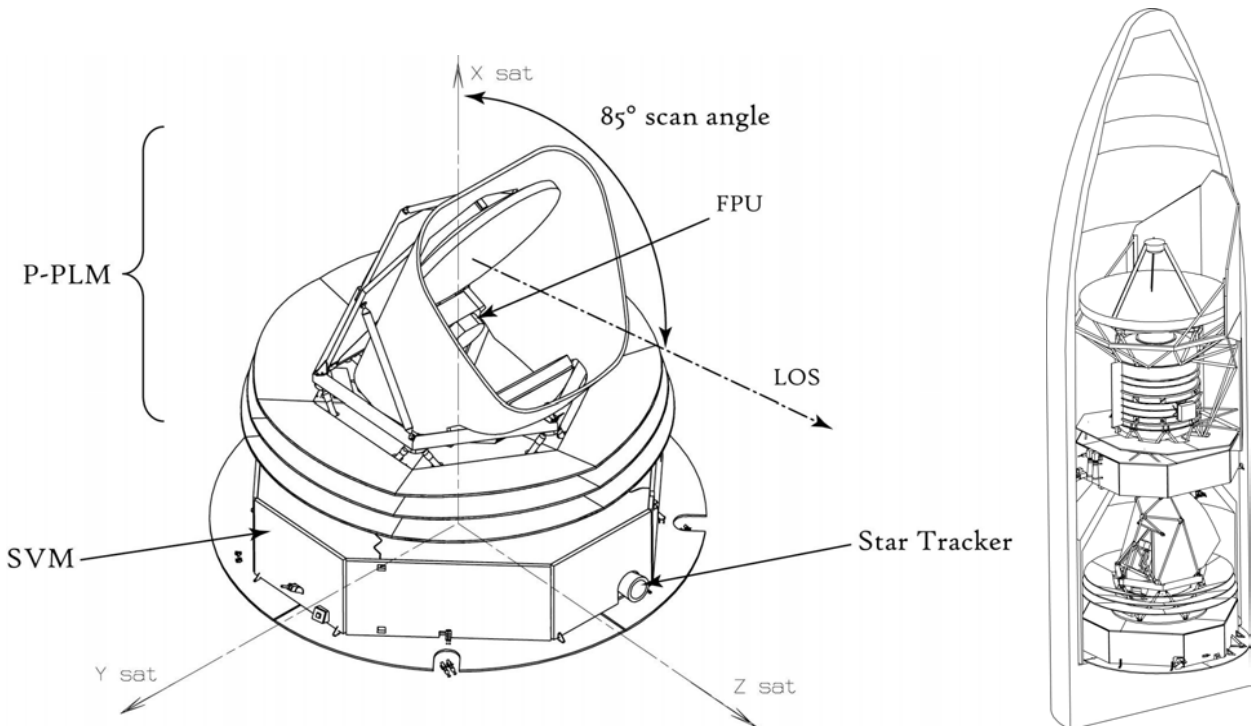

Fig. 1. Sketch of the PLANCK Satellite (left) and inside the Ariane V fairing together with HERSCHEL (right)

The satellite is scheduled for a launch in the first quarter of 2007 together with HERSCHEL with an ARIANE 5 launch vehicle from Kourou. The HERSCHEL/PLANCK spacecraft will acquire its nominal orbital position at 1.5 million Km from the earth, in an orbit around the second Lagrangian point of the Earth-Sun System. Once separated, HERSCHEL and PLANCK will perform independently their observing programmes.

The satellite will spin at 1 rpm around the "*X sat*", the "*spin axis*" (see Fig.1.), that will be always pointing in the satellite-sun direction. The telescope line of sight (LOS) is offset at 85° from the spin axis and will scan the sky in great circles, allowing both instruments to cover twice the entire sky in one year of continuous observations. Each pixel on the sky will be measured several time by the different detectors at each frequency and on many different time scales. This scanning strategy permits to reach the temperature stability and excellent control of systematic effects required by PLANCK's high precision measurements. The mapping scheme also allows good characterization and subtraction of any residual instrumental offsets and drifts. The multifrequency nature of the measurement provides for separation of astrophysical foregrounds from the cosmological signal.

**Satellite Description**

A high-precision CMB anisotropy measurement requires both extremely good sensitivity and rejection of systematic effects. The designs of the spacecraft, the telescope and the two instruments are driven by these two requirements.
The spacecraft follows a modular design concept. It is constituted of two modules: the service module (SVM) which provides all the servicing functions to the payload and also accommodates the warm instrument electronics; the payload module (P–PLM) which includes three thermal shields (V–grooves), the telescope and related structure and baffle, and the two instruments cold units. The three V–grooves, passively cooled at around 150K, 100K, and 60K respectively, will provide the needed thermal environment to allow the cooling of the two instruments (20K and 0.1K for LFI and HFI respectively). LFI will be cooled at 20K by an Hydrogen Sorption Cooler [3]. This novel vibration-less refrigerator works as a closed cycle cooler using Joule-Thomson (J-T) expansion of high pressure hydrogen gas and a set of six compressor elements which provide continuous circulation of hydrogen. The compressors are located on the SVM and connected to radiators for heat rejection. The gas is transported to the cold–end by pipes which are thermally interfaced to the V–grooves. On the cold–end the gas is liquefied to directly refrigerate the LFI. The Sorption Cooler will also provide a pre-cooling stage for the HFI 4K closed-cycle helium J–T cooler. The operating temperature of the HFI bolometers (0.1K) will be reached by a Benoit style open cycle dilution cooler. The two instruments share the focal region of the telescope: HFI in the centre and LFI in a ring around it; LFI provide also the mechanical interface between both instruments and the satellite.

**The PLANCK Telescope**

The design of the PLANCK telescope is probably the most advanced optical design ever conceived for accommodating large multi beam and multi-frequency focal plane units [4].

The optical design has been carried out by the Industrial Contractor, the Instrument teams, and ESA, within an activity aimed at optimising the telescope in the framework of the Payload Architect Study during the year 2000 [5]. The final design has been selected among several candidates as the best design in terms of wave front error over the whole field of view, straylight rejection, beam symmetry, angular resolution, and flatness of the focal surface. The PLANCK telescope represents a challenge for telescope technology and optical design for two reasons. First, both the HFI and LFI instruments require high optical performances over a focal region as large as 400 x 400 mm: high symmetry of the main beam up to 5° away from optical axis, excellent straylight rejection, high mirror reflectivity (~ 99 % of the whole telescope), very wide frequency coverage from 25 GHz to 1000 GHz. Second, the cryogenic environment (40 – 65 K) in which the telescope will operate will require an advanced manufacturing technology – for light weight optics – to avoid any thermal shocks and mirror deformations during the cool down. All these characteristics have never been obtained before in similar experiments.

The optical design is based on a two mirror off–axis scheme which offers the advantage of an unblocked aperture that maintains the diffraction by the secondary mirror and struts at very low levels. As in the Gregorian Aplanatic design, both the primary and the secondary mirrors have an ellipsoidal shape [6]. Unlike conventional Aplanatic designs, the PLANCK subreflector revolution axis is tilted with respect to the main reflector revolution axis, by an angle of 10.1°. The physical aperture of the mirrors is approximately 150cm x 190cm and 100cm x 100cm, for the primary and secondary reflector respectively. These give a projected aperture of 1.5 meter and an unblocked field of view as large as +/– 5° centered on the line of sight (LOS) of the telescope. The secondary mirror has been oversized in order to avoid under illumination of the primary for the off–axis feeds.

In order to reduce the straylight, a baffle will be installed around the telescope. The baffle acts also as a thermal shield to maintain at about 50K the physical temperature of the telescope itself.

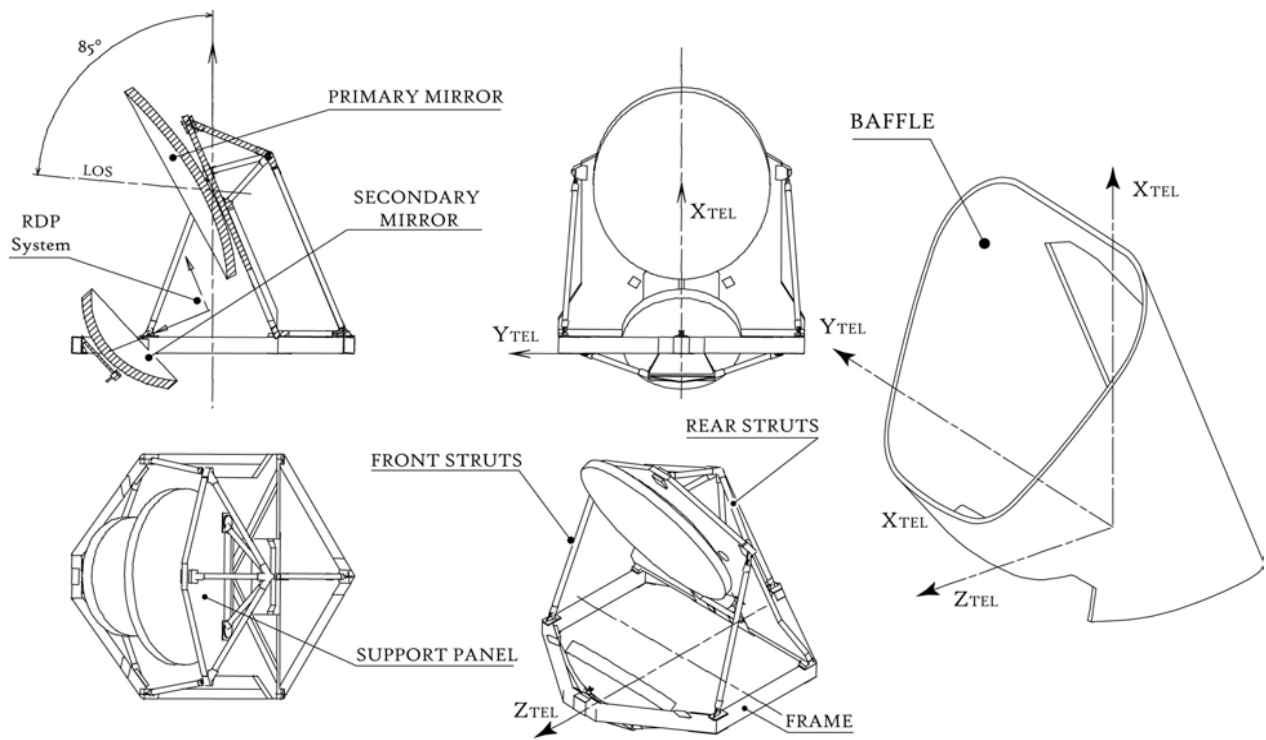

Fig. 2. Overall design of the PLANCK Telescope, structure and baffle (on the right). The baffle design is not in scale.

# THE LOW FREQUENCY INSTRUMENT

The Low Frequency Instrument (LFI) has been proposed by an International Consortium lead by IASF/CNR – Sezione di Bologna (formerly TESRE) since the Phase A study in 1996. Following NASA's COBE/DMR and MAP satellites, LFI is the third generation of radiometric instruments to image the CMB anisotropies. The instrument will map the whole sky at 30, 44, 70, and 100 GHz with a sensitivity per pixel of $\Delta T/T \sim (2 \div 7)$ µK and an angular resolution between 36' to 12' (33' to 10' as a goal) depending on frequency. LFI is intrinsically sensitive to the polarisation signal. In addition, in order to maximise the information from linear polarization of the incoming radiation, the LFI detectors have been optimized in orientation in order to observe the same spot on the sky during the same scan along the great circles with pairs of horns and Ortho Mode Transducers (OMTs) rotated by 45° in the sky [7].

The LFI is composed by an array of 46 state-of-the-art radiometers based on Indium Phosphide (InP) cryogenic HEMT amplifiers cooled at 20K in order to achieve the best possible sensitivity. To minimize the power dissipation in the focal area the LFI radiometers are split into two sub-assemblies (Front–End Unit and Back–End Unit) one located at the telescope focus, the other on the PLANCK service module at ~300K. The two parts are connected with a bundle of 92 rectangular waveguides which are designed to provide the necessary conductive thermal break between the cold and the warm parts of the instrument, while maintaining low insertion loss performance.

The twenty–three front end modules are located in a ring on the FPU as shown in the right panel of Fig. 2. Each module (see Fig. 3) contains the front end stage of two pseudo-correlation receivers both connected to one feed horn by an OMT. The OMT splits the signal entering into the horn in two linear orthogonal polarization components. In order to get linear polarization information of the incoming radiation, the detectors of LFI have been optimized in orientation. The number of modules has been chosen in order to roughly match the sensitivity per equal-size pixels on the sky for each frequency channel. In this way, LFI is composed by 2 modules at 30 GHz, 3 modules at 44 GHz, 6 modules at 70 GHz and finally 12 modules at 100 GHz, and each module is constituted by two independent radiometers. The radiometer design has been chosen to minimize the 1/f noise [8] by performing a continuous comparison between the sky signal and the signal coming from a internal blackbody load cooled at 4K [9]. InP HEMTs in cascaded gain stages are used since they exhibit the best noise performances in the LFI frequency range. The amplifiers at 30 and 44 GHz will use discrete InP HEMTs incorporated into a microwave integrated circuit (MIC). At these frequencies, cryogenic MIC amplifiers have demonstrated noise temperatures better than 10 K, with 20 % bandwidth. At 70 and 100 GHz LFI will use MMICs (Monolithic Microwave Integrated Circuits), which incorporate all circuit elements and the HEMT transistors on a single InP chip. Cryogenic MMIC amplifiers have been demonstrated at 75-115 GHz which exhibit ≤ 50 K noise from 90-105 GHz. The LFI will fully exploit both MIC and MMIC technologies at their best.

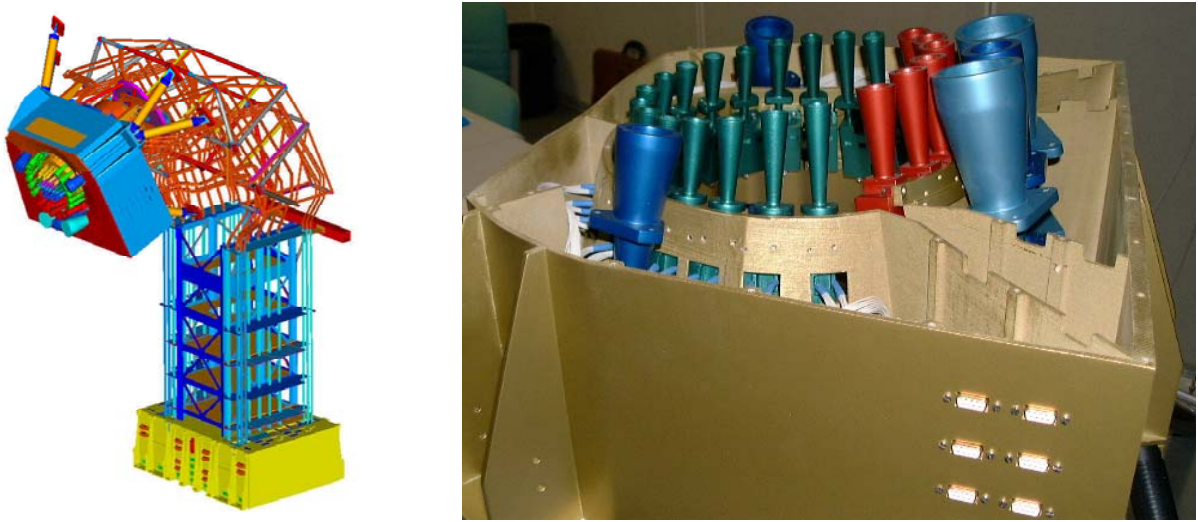

Fig. 2. The entire Low Frequency Instrument (left) and a mock-up of the 20K Focal Plane Unit (right)

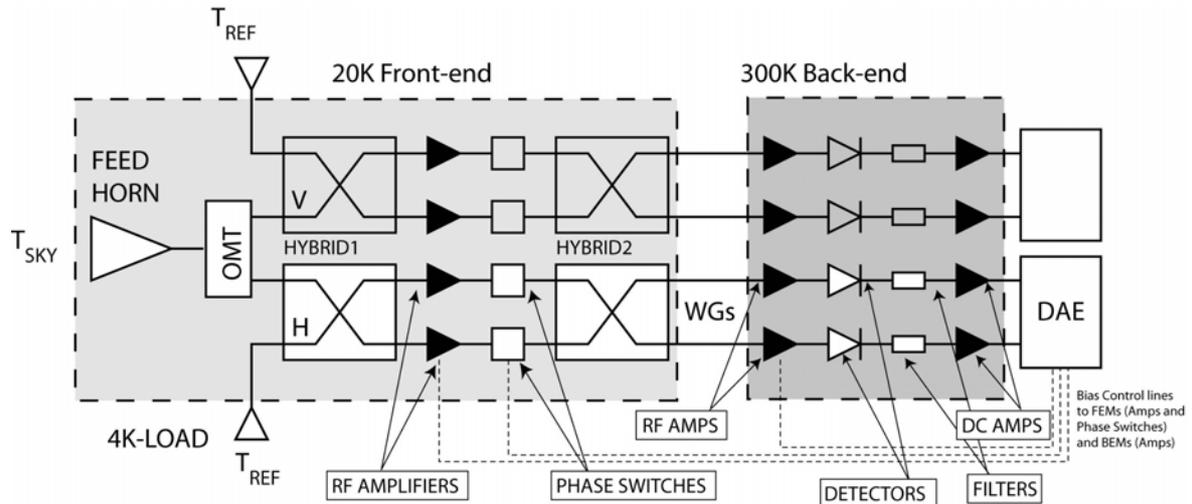

Fig. 3. The Receiver module of LFI which includes two radiometer chains each connected at one output of the OMT.

The 32 back end modules include additional RF amplifiers, detectors, filters, and DC amplifiers as shown in Fig. 3. At the output of each radiometer channel the sky signal and the reference signal are detected by the square law diodes pair alternatively and synchronously with the phase switches located in the front end. For each module, a set of 4 rectangular waveguides, approximately 2 meter long, connects the front end and the back end sections. Each waveguide is twisted and bent in different planes and with different angles in order to fit into the Focal Plane Unit and to allow the integration of the LFI and HFI instruments. To maintain low insertion loss, low return loss and at the same time provide an adequate thermal break, the waveguides are manufactured in sections with different materials. An electroformed copper waveguide section (which presents the abovementioned bends and twists) is attached to the Front-End module. This section will be attached with custom flanges to a Stainless Steel straight section which will act as a thermal break. Part of the stainless steel section is gold plated inside in order to increase the electrical conductivity. The whole instrument, including the waveguide support structure is shown in the left panel of Fig. 2.

**THE LFI OPTICAL INTERFACES**

As previously mentioned, the radiation coming from the sky is focalized by the telescope and captured by the feed horns located in the focal region. The coupling between the feeds and the external environment constitutes the Optical Interfaces. Ideally, the external environment is composed by the PLANCK telescope only. In practice, because of the imperfect feed response and coupling between the feed and telescope, the environment is constituted of all the satellite surfaces (including the telescope itself) and celestial objects which can emit, reflect, or scatter, radiation within the operational bandwidth of the LFI detectors.

The optical interfaces have a potentially strong effect on the LFI scientific capabilities because are related to the angular resolution and to the level of the straylight. In fact, the objective of the optical design optimisation is to obtain the best angular resolution achievable compatible with the very stringent (at μK level) straylight requirements. Both angular resolution on the sky and the level of straylight contamination depend on the edge taper, i.e. on the feed horn aperture diameter. The more the illumination of the telescope is peaked, the more the straylight is low and the more the angular resolution is poor. On the contrary, a strong illumination of the primary mirror increases the angular resolution but also increases the side lobe levels. The optical interfaces are optimized essentially by an appropriate design of the feed horns. Dual profiled corrugated horns have been selected as the best design in terms of shape of the main lobe and level of the side lobes, control of the phase centre position, and compactness. LFI dual profiled horns are composed by an exponential profile near the aperture attached to a "Sine-Squared" profile section [10]. The shape of the two sections can be used to control, as much as possible, the position and frequency stability of the phase centre and the compactness of the structure. The design of the feed horns starts from the electromagnetic requirements (return loss, edge taper, side lobe level, cross–polar level) and the mechanical requirements (phase centre position, maximum envelope, length of the corrugation section). Because of the off–axis position of the LFI feeds (typically at 150 mm away from the centre) and the asymmetric response of the telescope, the requirements are different for different horns even at the same frequency [11].

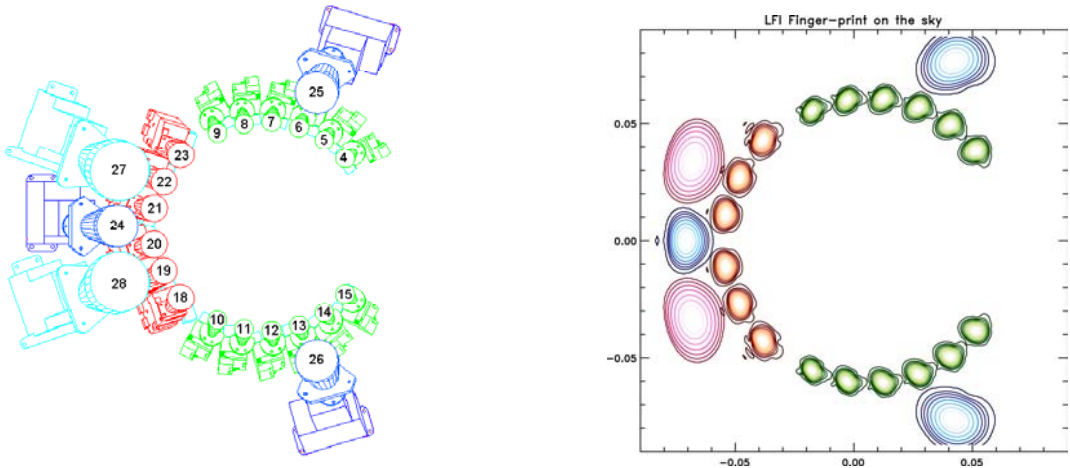

Fig. 4. Left: Layout of the LFI focal plane. The horns number 4÷15 are at 100 GHz; the horns 18÷23 are at 70 GHz. The numbers 24÷26 are at 44 GHz. The 27 and 28 are at 30 GHz. Right: the corresponding main beams on the sky. The scan direction is toward the vertical.

Because of the size of the corrugations, different manufacturing techniques have been selected depending on frequency: electroerosion for 30 GHz and 44 GHz horns and electroformation for the higher frequency horns (70 GHz and 100 GHz). Prototypes have been fabricated within the development activity of LFI. Feed horn models have been designed, manufactured and tested at 30, 70 and 100 GHz.

In order to optimize the LFI optical interfaces, a dedicated activity has been setup since '98. The main beam aberrations have been studied extensively [12][13]. The beam distortions produce a degradation in terms of angular resolution and add systematic noise on the pixels. Several studies have been carried out to evaluate the straylight for PLANCK (e.g. [14]). One of the main difficulties is to calculate with high accuracy the radiation pattern of the feed–telescope–baffle system over the whole solid angle. A work devoted to refine the edge taper values which optimize the angular resolution maintaining the straylight at acceptable level, is currently in progress.

**CONCLUSIONS**

The PLANCK Low Frequency Instrument is one of the two instruments onboard the ESA PLANCK satellite. It is designed to obtain high resolution, high sensitivity, full-sky imaging of the Cosmic Microwave Background including polarization at frequencies between 30 GHz and 100 GHz. LFI will use state of the art HEMT–based radiometers and it is designed to minimize systematic effects. The PLANCK telescope and the corrugated LFI feed horns represent one of the most advanced systems for mm–wave optics devoted to experimental cosmology. The multi–beam, multi–frequency system of LFI allows the measurement of the sky signal with unprecedented sensitivity and angular resolution at these frequencies.

In the framework of PLANCK, a strong effort has been dedicated to the optimisation of the coupling between the instruments and the telescope thanks to the close interaction between the instrument teams, ESA, and the Industry. This fruitful synergy is continuously assured by dedicated working group activities with the aim to improve the knowledge of the entire optical system at the best, with accurate simulations and measurements. Several telescope related activities are planned in the next future. The accurate prediction and verification of the telescope performance at cryogenic temperature, the procedures and verification of the telescope and instruments alignment, the refinement of the feed/telescope system full pattern simulations, are only few examples. Because of the complexity of the PLANCK optical system and of the instruments optical interfaces, these studies will provide a sensible improvement on the design and technology of microwave and mm–wave range antennas.

**ACKNOWLEDGMENTS**

This paper represents the work carried out by a large number of people who are working on the development of the PLANCK mission. We wish to thank the Project and the Instrument Consortia.

**REFERENCES**

[1] N. Mandolesi, et al., "The Low Frequency Instrument", *response to the Announcement of Opportunity for the FIRST/Planck Programme,* February 1998.

[2] J-L. Puget et al., "The High Frequency Instrument", *response to the Announcement of Opportunity for the FIRST/Planck Programme,* February 1998.

[3] L A Wade, P Bhandari, R C Bowman, C Paine, G Morgante, A Lindensmith, D Crumb, M Prina, R Sugimura, D Rapp, 2000, Advances in Cryogenic Eng., Vol. 45, pp.499-506, 2000.

[4] D. Dubruel, M. Cornut, Fargant, T. Passvogel, P. De Maagt, et al., "Very Wide Band Antenna Design for Planck Telescope Project Using Optical and Radio Frequency Techniques" in *Millennium Conference on Antennas & Propagation - AP2000, edited by Danesy, D. & Sawaya, H., ESA Conference Proceedings SP-444, European Space Agency,* CD-ROM, 2000.

[5] F. Villa, M. Bersanelli, C. Burigana, R.C. Butler, N. Mandolesi, et al., "The Planck Telescope", in *Experimental Cosmology at Millimetre Wavelengths - 2K1BC Workshop, AIP Conference Proceedings,* vol. 616, pp. 224-228, 2002.

[6] F. Villa, N. Mandolesi, C. Burigana, "A Note on the Planck Aplanatic Telescope", *Int. Rep. TeSRE/CNR,* 221/1998, September 1998.

[7] F. Villa, N. Mandolesi, M. Bersanelli, R.C. Butler, C. Burigana, et al., in *Astrophysical Polarized Backgrounds, APB Workshop, AIP Conference Proceedings,* vol. 609, pp. 144-149, 2002.

[8] L. Valenziano, M. Bersanelli, R.C. Butler, F. Cuttaia, N. Mandolesi, et al., "The 4K Reference Load for the Planck Low Frequency Instrument", in *Experimental Cosmology at Millimetre Wavelengths - 2K1BC Workshop, AIP Conference Proceedings,* vol. 616, pp. 219-223, 2002.

[9] M. Seiffert, A. Mennella, C. Burigana, N. Mandolesi, M. Bersanelli, et al., "1/f Noise and other Systematic Effects in the Planck-LFI Radiometers", A&A, 2002, in press.

[10] G.G. Gentili, R. Nesti, G. Pelosi, V. Natale, "Compact dual-profiled Corrugated Circular Waveguide Horns", *Electronics Letters,* Vol. 36, pp. 486-487, March 2000.

[11] M. Sandri, M. Bersanelli, C. Burigana, R.C. Butler, M. Malaspina, et al., "Planck Low Frequency Instrument: Beam Patterns", in *Experimental Cosmology at Millimetre Wavelengths - 2K1BC Workshop, AIP Conference Proceedings,* vol. 616, pp. 242-244, 2002.

[12] C. Burigana, D. Maino, N. Mandolesi, E. Pierpaoli, M. Bersanelli, et al., "Beam distortion effects on anisotropy measurements of the cosmic microwave background", *Astron. Astrophys. Suppl. Ser.*, Vol. 130, pp. 551-560, June 1998.

[13] N. Mandolesi, M. Bersanelli, C. Burigana, K.M. Górski, E. Hivon, et al., "On the performance of Planck-like telescopes versus mirror aperture", *Astron. Astrophys. Suppl. Ser.*, Vol. 145, pp. 323-340, August 2000.

[14] C. Burigana, D. Maino, K.M. Górski, N. Mandolesi, M. Bersanelli, et al., "PLANCK LFI: Comparison between Galaxy Straylight Contamination and other systematic effects", *Astron. & Astrophys.,* Vol. 373, pp. 345-358, 2001.